\documentstyle[prb,aps]{revtex}

\newcommand{\beq}{\begin{equation}}
\newcommand{\eeq}{\end{equation}}
\newcommand{\bqn}{\begin{eqnarray}}
\newcommand{\eqn}{\end{eqnarray}}
\newcommand{\bqns}{\begin{eqnarray*}}
\newcommand{\eqns}{\end{eqnarray*}}
\newcommand{\bary}{\begin{array}}
\newcommand{\eary}{\end{array}}
\newcommand{\non}{\nonumber}

\begin{document}
\draft

\twocolumn[\hsize\textwidth\columnwidth\hsize\csname %
@twocolumnfalse\endcsname
\title{Acoustic wave propagation in an one-dimensional layered system}
\author{Pi-Gang Luan and Zhen Ye}
\address{Wave Phenomena Laboratory, Department of Physics,
National Central University, Chung-li, Taiwan 320}
\date{\today}
\maketitle
\begin{abstract}
Propagation of acoustic waves in an one-dimensional water duct
containing many air filled blocks is studied by the transfer
matrix formalism. Energy distribution and interface vibration of
the air blocks are computed. For periodic arrangement band
structure and transmission rate are calculated analytically,
whereas the Lyapunov exponent and its variance are computed
numerically for random situations. A distinct collective behavior
for localized waves is found. The results are also compared with
optical situations.
\vspace{1cm}
\end{abstract}
 ]
\section{Introduction}

Propagation of waves in periodic and disordered media has
been and continues to be an interesting subject for physicists
\cite{Sheng,Ish,Hodges83,Baluni,Matsuda,SDE,Sor,Maradudin93,Maradudin95,Ye1,Ye2,Sch,Frank,Cha,Deych,Ye}.
When propagating in media with inhomogeneities, waves are subject
to multiple scattering, which leads to many peculiar phenomena
such as band structures in periodic media and wave localization in
random media\cite{Leung,Yab,Kush,ES,San,Tor,Anderson,TAA}.

The propagation of waves in one dimensional (1D) systems has
attracted particular interest from scientists because in higher
dimensions the interaction between waves and scatterers is so
complicated that the theoretical computation is rather involved
and most solutions require a series of approximations which are
not always justified, making it difficult to relate theoretical
predictions to experimental observations. Yet wave localization in
one dimension (1D) poses a more manageable problem which can be
tackled in an exact manner by the transfer matrix method
\cite{Hodges83,Baluni,Matsuda,SDE,Sor,Maradudin93,Maradudin95,Deych}.
Moreover, results from 1D can provide insight to the problem of
wave localization in general and are suitable for testing various
ideas. Indeed, over the past decades considerable progress has
been made in understanding the localization behavior in 1D disordered
systems\cite{Hodges83,Baluni,Matsuda,SDE,Sor,Maradudin93,Maradudin95,Deych}.
However, a number of important issues remained
untouched. These issues include, for example, how waves are
localized inside the media and whether there is a distinct feature
for wave localization which would allow to differentiate the
localization from residual absorption effect without
ambiguity\cite{Frank,Cha}. Results from the statistical analysis
of the scaling behavior in 1D random media is not conclusive.
Another question could be whether the localized state is a phase
state which would accommodate a more systematic
interpretation\cite{Umezawa}. Furthermore, the study of acoustic
propagation in 1D random media is relatively scarce. All these
motivate us to consider wave propagation in 1D media further, with
emphasis on the acoustic wave propagation.

In this paper we study the problem of acoustic wave propagation in
an one dimensional water duct containing many air blocks either
regularly or randomly but on average regularly distributed inside
the duct. The frequency band structures and wave transmission are
computed numerically. We show that while our results affirm the
previous claim that all waves are localized inside an 1D medium
with any amount of disorder, there are, however, a few distinctive
features in our results. Among them, in contrast to optical
case\cite{Deych}, there is no universal scaling behavior in the
present system. In addition, when waves are localized, a
collective behaviour of the system emerges. We will also show the
energy distribution in the water duct.

This paper is organized as follows. The next section we explain
the the model employed, discuss the transfer matrix method and
derive relevant formulas. In section III numerical results and
discussions are given. We then summarize the paper in section IV.

\section{Model and method}

\subsection{System setup}

We study the system consisting air blocks inside a water duct. The
system is chosen because air filled blocks are strong acoustic
scatterers. This can be seen as follows. The scattering is largely
controlled by the acoustic impedance which is defined as $\rho c$,
with $\rho$ and $c$ being the mass density of the medium and the
acoustic phase speed respectively. The acoustic impedance ratio
between water and air is about 3500. This large contrast leads to
strong scattering, making the system of air blocks in water an
ideal candidate for the study of acoustic scattering.

\begin{figure}[hbt]
\begin{center}
\epsfxsize=3in \epsffile{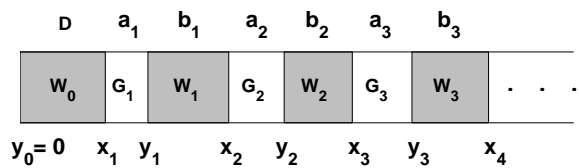}\vspace{5mm}
\caption{\label{figure1}\small Regular and random arrangements of
air blocks in the duct. The symbols ${\mathbf G}_{j}$, ${\mathbf
W}_{j}$ denote the air and water layers, $a_{j}$ and $b_{j}$
represent their thicknesses, the $j$-th air block and the water
separated by the $j$-th and $j+1$-th air blocks, denoted as the
$j$-th water block, is regarded as one unit and is labeled as the
$j$-th unit; therefore $d_{j}=a_{j}+b_{j}$ represents the
thickness of the $j$-th unit. $x_{j}=D+\sum^{j-1}_{l=1}d_{l}$ and
$y_{j}=x_{j}+a_{j}$ represent the pisitions of air-water and
water-air interfaces. For regular arrangement $a_{j}=a$,
$b_{j}=b$, $d_{j}=d=a+b$ and $x_{j}=D+(j-1)d$, $y_{j}=x_{j}+a$.}
\end{center}
\end{figure}

The 1D acoustic system we consider is illustrated by
Fig.~\ref{figure1}. Assume that $N$ air blocks of thickness
$a_{j}$ ($j=1,\dots, N$) are placed regularly or randomly in a
water duct with length $L$ measured from the left boundary of the
duct (LB). The distance from LB to the left interface of the first
air block is $D$. For simplicity while without compromising
generality, in later numerical computation we set the thickness of
each air block to the same value $a$. The air fraction is clearly
$\beta = Na/L$, the average distance between two adjacent air or
water blocks is $\langle d\rangle = L/N=a/\beta$, and the average
thickness of water blocks is $\langle
b\rangle=(D+\sum^{N-1}_{j=1}b_{j})/N$. The degree of randomness
for the system is controlled by a parameter $\Delta$ in such a way
that the thickness of the $j$-th water block is $b_{j}=\langle
b\rangle(1+\delta_{j})$ with $\delta_{j}$ being a random number
within the interval $[-\Delta,\Delta]$; the regular case
corresponds to $\Delta =0$. An acoustic source placed at LB
generates monochromatic waves with an oscillation
$v(t)=ve^{-i\omega t}$. Transmitted waves propagate through the
$N$ air blocks and travel to the right infinity. In order to avoid
unnecessary confusion, possible effects from surface tension,
viscosity or any absorption are neglected.

For convenience, we use the dimensionless quantity $k\langle
b\rangle$ to measure the frequency, where $k=\omega/c$ is the wave
number in water blocks. Similarly, $k_{g}$ represents the wave
number in air blocks. We also define the following parameters for
later use \beq g=\frac{\rho_{g}}{\rho},\;\;
h=\frac{c_{g}}{c}=\frac{k}{k_{g}},\;\; q^{2}=gh,\;\;\eta=\ln q.
\eeq

\subsection{Wave propagation and state vector}

We use the transfer matrix method\cite{Baluni} for solving the
wave propagation in the system. Dropping out the time factor
$e^{-i\omega t}$, the wave propagation obeys Helmholtz equation
\beq p_{m}''(x)+k^{2}_{m} p_{m}(x)=0,\label{helm} \eeq in which
$p_{m}(x)$ is the pressure field, and the subscript $m$ refers to
the medium that can be either water or air, depending on where $x$
is located. Within any layer (air blocks or water blocks), the
wave is \beq
p_{m}(x)=A_{m}e^{ik_mx}+B_{m}e^{-ik_mx}.\label{pab} \eeq where
$A_{m}e^{ik_mx}$ represents the wave transmitted away from the
source to the right and $B_{m}e^{-ik_mx}$ the wave reflected
towards the source. In terms of $A_m$ and $B_m$, the corresponding
velocity field $u_{m}$, which is another dynamical variable
describing the oscillation of the medium, can be calculated as
\bqn {u}_{m}(x) &=&\frac{1}{i\omega\rho_{m}} p'_{m}(x)\non\\
&=&\frac{1}{\rho_m c_m}\left [A_{m}e^{ik_mx}-B_{m}e^{-ik_mx}\right
],\label{pu} \eqn where $\rho_{m}$ refers to the mass density of
medium $m$.

Define a state vector \beq {\mathbf
S}_{m}(x)\equiv\left(\begin{array}{c}S^1\\S^2\end{array}\right) =
\left(\bary{c}A_m (x)\\
\\B_m (x)\eary\right)=
\left(\bary{c}A_m e^{i k_{m} x}\\
\\B_m e^{-i k_{m} x}\eary\right),\label{state}
\eeq then $p_{m}(x)$ and $u_{m}(x)$ can be determined by ${\mathbf
S}_{m}(x)$: \beq p_m = S^1 +S^2, \ \ \ u_m = \frac{1}{\rho_m
c_m}(S^1-S^2).\eeq We denote the state vector in the $j$-th air
block as ${\mathbf G}_{j}(x)$ with two components
$G^{1}_{j}(x)=G^{1}_{j}e^{ik_{g}x}$ and
$G^{2}_{j}(x)=G^{2}_{j}e^{-ik_{g}x}$ and in the $j$-th water block as
${\mathbf W}_{j}(x)$ with $W^{1}_{j}(x)=W^{1}_{j}e^{ikx}$ and
$W^{2}_{j}(x)=W^{2}_{j}e^{-ikx}$.

By invoking the condition that the
pressure and velocity fields are continuous across the interfaces
separating water and air, one finds
\[
{\mathbf W}_{j-1}(x_{j})=J{\mathbf G}_{j}(x_{j}),\;\;\;
{\mathbf G}_{j}(y_{j})=J^{-1}{\mathbf W}_{j}(y_{j}),
\]
with
\beq J=q^{-1}\left(\bary{cc}\cosh\eta&\sinh\eta\\
\sinh\eta&\cosh\eta\eary\right).\label{wgj} \eeq

For  wave propagation in the $j$-th air or water layer, we have
\beq {\mathbf G}_{j}(x_{j})=U_{g}(a_{j}){\mathbf G}_{j}(y_{j}),
\label{UU1}\eeq with $$ U_{g}(a_{j})
=\left(\bary{cc}e^{-ik_{g}a_{j}}&0\\
0&e^{ik_{g}a_{j}}\eary\right), $$ and \beq {\mathbf W}_{j}(y_{j})=
U(b_{j}){\mathbf W}_{j}(x_{j+1}), \label{UU2}\eeq with $$
U(b_{j})=\left(\bary{cc}e^{-ikb_{j}}&0\\
0&e^{ikb_{j}}\eary\right).$$ From these results the transfer
matrix $M_{j}$ for $j$-th unit is \beq
M_{j}=JU_{g}(a_{j})J^{-1}U(b_{j}).\label{juju} \eeq and the state
vectors in water blocks satisfies \beq {\mathbf
W}_{j-1}(x_{j})=M_{j}{\mathbf W}_{j}(x_{j+1}). \eeq Therefore any
two state vectors of the water blocks are connected as \beq
{\mathbf W}_{j_1-1}(x_{j_1})=M_{j_1,j_2}{\mathbf
W}_{j_2}(x_{j_2+1}),\label{wmw} \eeq where \beq
M_{j_{1},j_{2}}=M_{j_{1}}M_{j_{1}+1}\cdots M_{j_{2}}, \;\;\;1\leq
j_{1}\leq j_{2}\leq N. \label{mmm} \eeq From Eq.(\ref{juju}) and
(\ref{mmm}) one can easily prove that all $M_{j_{1},j_{2}}$ are
uni-modular matrices, a result of energy conservation. We denote
\beq M_{1,N} \equiv {\cal M} = \left(\begin{array}{cc} {\cal
M}_{11} & {\cal M}_{12}\\ {\cal M}_{21} & {\cal
M}_{22}\end{array}\right).\eeq A simple reduction leads to \beq
{\cal M}_{21} = {\cal M}_{12}^*, \ \ \ {\cal M}_{22} = {\cal
M}_{11}^*.\eeq It is clear that the transfer matrix $M_{1,N}$
connects the first water block and the last water block.

Imposing the boundary conditions at LB \beq v=\frac{1}{\rho
c}[W^{1}_{0}-W^{2}_{0}], \label{bdc1}\eeq and \beq
W^{2}_{N}=0,\label{bdc2} \eeq we obtain \bqn {\mathbf
W}_{0}(x_{1})&=&\frac{\rho cv}{{\cal M}_{11}\,e^{-ikD}-{\cal
M}^{*}_{12}\,e^{ikD}} \left(\bary{c}{\cal M}_{11}\\ \\ {\cal
M}^*_{12}\eary\right). \eqn The condition in (\ref{bdc1}) relates
${\mathbf W}_{0}$ to the LB oscillation $v$, and the second
relation (\ref{bdc2}) results from the fact that there is no
reflection at the rightmost boundary in Fig.~1.

\subsection{Parameterization for regular system}

In the situation of regular arrangement of air blocks, $a_{j}=a$,
$b_{j}=b$, $d_{j}=d$, $M_{j}=M$ and ${\cal M}=M_{1,N}=M^{N}$.
According to Eq.(\ref{juju}), $M$ acquires the form \beq
M=\left(\bary{cc}\alpha&\beta\\ \beta^{*}&\alpha^{*} \eary\right)
\eeq with \beq \mbox{det}(M) = 1, \label{apbt} \eeq where \beq
\bary{l}\alpha = e^{-ikb}[\cos k_{g}a-i\sin k_{g}a\cosh
2\eta],\non \\ \\ \beta = ie^{ikb}\sin k_{g}a\sinh
2\eta,\eary\label{ab} \eeq are functions of three parameters:
$k_{g}a$, $kb$ and $\eta$. In order to express physical quantities
in a simple closed form, we use the parametrization given in
\cite{Matsuda}. We define
\[
\mbox{tr}(M)\equiv 2\cos\theta,
\]
then the eigenvalue equation $\lambda^{2}-\mbox{tr}(M)\lambda+1=0$
yields two eigenvalues $\lambda=e^{\pm i \theta}$. Since
$\mbox{tr}(M) =\alpha+\alpha^{*}=2\alpha_{R}$, and thus $\alpha_R
= \cos\theta$, Eq. (\ref{apbt}) becomes
\[
\alpha^{2}_{I}-|\beta|^{2}=\sin^{2}\theta,
\]
where $\alpha_{R}=\mbox{Re}(\alpha)$ and $\alpha_{I}=\mbox{Im}(\alpha)$
represent the real and imaginary parts of $\alpha$, respectively.

Following \cite{Matsuda}, we adapt the parametrization, \beq
\bary{l}\alpha(\theta,\chi) \equiv
\cos\theta-i\sin\theta\,\cosh2\chi,\non\\ \\
\beta(\theta,\chi,\delta) \equiv
i\,e^{i\delta}\sin\theta\,\sinh2\chi,\eary \label{abeta} \eeq
where $\theta$ and $\chi$ in general are complex variables and
$\delta$ is chosen as $kb$. Comparing Eq. (\ref{ab}) and
(\ref{abeta}), the following relations are established: \bqn
\cos\theta && = \cos\!k_g a\,\cos\!kb-\sin\!k_g
a\,\sin\!kb\,\cosh2\eta,\non\\ \sin\theta\,\cosh2\chi && =
\cos\!k_g a\,\sin\!kb +\sin\!k_g a\,\cos\!kb\,\cosh2\eta,\non\\
\sin\theta\,\sinh2\chi &&= \sin\!k_g a\,\sinh2\eta.\label{sinsinh}
\eqn From these relations one finds that $\cos\theta$,
$\sin\theta\cosh 2\chi$ and $\sin\theta\sinh 2\chi$ are all real,
providing restrictions on the possible values of $\theta$ and
$\chi$. Furthermore, using the definition (\ref{abeta}), it is
straightforward to verify that \bqn [M(\theta,\chi,\delta)]^n & =&
M(n\theta,\chi,\delta)\non\\
&=&\left(\bary{cc}\alpha(n\theta,\chi)&
\beta(n\theta,\chi,\delta)\\
\beta^{*}(n\theta,\chi,\delta)&\alpha^{*}(n\theta,\chi)\eary\right),\label{mn}
\eqn where $n$ is an integer.

In this parametrization, matrix $M$ becomes a function of
$\theta,\chi,$ and $\delta$. A direct computation shows that the
two eigen-vectors of matrix $M$ can be written as \beq {\mathbf
X}_{1}=\left(\bary{cc}\cosh\chi \\
e^{-i\delta}\sinh\chi\eary\right),\ \ \  {\mathbf
X}_{2}=\left(\bary{cc}e^{i\delta}\,\sinh\chi \\
\cosh\chi\eary\right), \eeq corresponding to the eigen-values
$e^{\pm i\theta}$ respectively.

\subsection{Determination of $p$ and $u$}

Denoting the pressure and velocity fields on left and right
boundaries of $j$-th air block as $p^{L}_{j}$, $p^{R}_{j}$ and
$u^{L}_{j}$, $u^{R}_{j}$, we have \bqn p^{L}_{j} &=&
W^1_{j-1}(x_{j})+W^2_{j-1}(x_{j}),\non\\ p^{R}_{j} &=&
G^1_{j}(y_{j})+G^2_{j}(y_{j}),\non\\ && \label{ppuuu}\\ u^{L}_{j}
&=&
\frac{1}{\rho\,c}[\,W^1_{j-1}(x_{j})-W^2_{j-1}(x_{j})\,],\non\\
u^{R}_{j} &=&
\frac{1}{\rho_{g}c_{g}}[\,G^1_{j}(y_{j})-G^2_{j}(y_{j})\,].
\non\eqn The wave vectors ${\mathbf W}_{j-1}(x_{j})$ and ${\mathbf
G}_{j}(y_{j})$ are determined by the following equations: \bqn
{\mathbf W}_{0}(D) &=& M_{1,j-1}{\mathbf W}_{j-1}(x_{j})\non\\ &=&
M_{1,j-1}JU_{g}(a_{j}){\mathbf G}_{j}(y_{j}). \label{ww}\eqn

For the case of regularly arranged air blocks, by Eqs.~(\ref{mn}),
(\ref{ppuuu}), and (\ref{ww}), $p^{L}_{j}$, $p^{R}_{j}$ and
$u^{L}_{j}$, $u^{R}_{j}$ are explicitly given by \bqn p^{L}_{j}
&=& \rho
cv\,\left[\frac{\alpha[(N-j+1)\theta]+\beta^{*}[(N-j+1)\theta]}
{\alpha(N\theta)\,e^{-ikD}-\beta^{*}(N\theta)\,e^{ikD}}\right],\non\\
p^{R}_{j} &=& \rho
cv\,\left[\frac{\alpha[(N-j)\theta]\,e^{-ikb}+\beta^{*}[(N-j)\theta]\,e^{ikb}}
{\alpha(N\theta)\,e^{-ikD}-\beta^{*}(N\theta)\,e^{ikD}}\right],\non\\
&& \hspace{2cm} \non \\ & & \label{ppuu}
\\ u^{L}_{j} &=&
v\,\left[\frac{\alpha[(N-j+1)\theta]-\beta^{*}[(N-j+1)\theta]}
{\alpha(N\theta)\,e^{-ikD}-\beta^{*}(N\theta)\,e^{ikD}}\right],\non\\
&& \hspace{2cm}\non\\ u^{R}_{j} &=&
v\,\left[\frac{\alpha[(N-j)\theta]\,e^{-ikb}-\beta^{*}[(N-j)\theta]\,e^{ikb}}
{\alpha(N\theta)\,e^{-ikD}-\beta^{*}(N\theta)\,e^{ikD}}\right].
\non \eqn

\subsection{Band structures}

According to the Bloch theorem, the eigen-modes of wave field $p$
and velocity field $u$ in an infinite periodic medium can be
written as \beq p(x)=\xi(x)e^{iKx},\;\;\;
u(x)=\zeta(x)e^{iKx}\label{pxuz} \eeq where $\xi(x)$ and
$\zeta(x)$ are periodic functions satisfying $\xi(x+d)=\xi(x)$,
$\zeta(x+d)=\zeta(x)$ and $K$ is the usual Bloch wave number.

Eq.~(\ref{pxuz}) implies \bqn e^{-iKd}{\mathbf
W}_{j}(x_{j}+d)&=&{\mathbf W}_{j-1}(x_{j})\nonumber\\
&=&M(\theta,\chi,\delta){\mathbf W}_{j}(x_{j}+d). \eqn Therefore
$e^{-iKd}$ is the eigenvalue of matrix $M$ and equals
$e^{i\theta}$. Substituting $Kd = \theta$ in Eq.~(\ref{sinsinh}),
we obtain the dispersion relation \beq \cos Kd=\cos k_{g}a\, \cos
kb -\cosh 2\eta\,\sin k_{g}a\, \sin kb,\label{disp}\eeq which
describe the band structure. When
\[
|\cos\!k_g a\,\cos\!kb-\sin\!k_g a\,\sin\!kb\,\cosh2\eta|\leq 1,
\]
the solution for $K$ is a real number. When
\[
|\cos\!k_g a\,\cos\!kb-\sin\!k_g a\,\sin\!kb\,\cosh2\eta|> 1,
\]
the solution for $K$ is complex and is of the form \beq
K=\frac{n\pi}{d}+iK_{I}, \eeq where $n$ is an integer in order to
satisfy Eq.~(\ref{disp}). The frequency ranges within which real
solutions for $K$ can be deduced define the pass bands, while the
ranges rending the complex solutions for $K$ determine the stop
bands (band gaps).

\subsection{Transmission, reflection, energy flow
and Lyapunov exponent}

Waves propagating through $N$ air blocks will be scattered many times
before they go out. The total transmission and reflection rates can be
obtained from scattering matrix $M^{s}$,
\[
 M^{s}=\left(\bary{cc}1/t& r^{*}/t^{*}\\ r/t&1/t^{*}
\eary\right),
\]
where $t$ and $r$ are transmission and reflection coeffcients, and
$T=|t|^2$ and $R=|r|^2$ define the transmission and reflection
rates. Based on the previous discussion \beq M^{s}=M_{1,N}={\cal
M} \eeq and \beq T_{N}=\frac{1}{|{\cal M}_{11}|^2},\;\,
R_{N}=\frac{|{\cal M}_{12}|^2}{|{\cal M}_{11}|^2}.\label{tr} \eeq
Note that in Eq. (\ref{tr}) the relation $T_{N}+R_{N}=1$ is
satisfied, which is a consequence of energy conservation since the
system we considered does not absorb any energy.

Another consequence of energy conservation is that energy flow
along the whole duct is a constant. Recall that the time averaged
energy flow ${\cal J}(x)$ is calculated as \beq {\cal J}(x) =
\frac{1}{2}\mbox{Re}\left[\;p(x)u^{*}(x)\right]. \eeq In the
presence of N air blocks, the flow can be easily obtained as \beq
{\cal J}_{N}=\frac{{|W^1_{N}|^{2}}}{2\rho c}=\frac{\rho
v^{2}}{2}\left[\, \frac{c}{|{\cal M}_{11}e^{-ikD}-{\cal
M}^{*}_{12}e^{ikD}|^{2}}\,\right]. \label{J} \eeq

For the case of regularly placed $N$ air blocks we can obtain \beq
T_{N}=\frac{1}{|\alpha(N\theta)|^2}
=\frac{1}{1+\sin^{2}N\theta\,\sinh^{2}2\chi},\label{regtn} \eeq
\beq R_{N}=\frac{|\beta(N\theta)|^2}{|\alpha(N\theta)|^2}
=\frac{\sin^{2}N\theta\,\sinh^{2}2\chi}
{1+\sin^{2}N\theta\,\sinh^{2}2\chi},\label{regrn} \eeq and \bqn
{\cal J}_{N} &=& \frac{\rho v^{2}}{2}
\left[\,\frac{c}{C+A\cos2N\theta+B\sin2N\theta}\,\right]\non\\ &=&
\frac{\rho v^{2}}{2}
\left[\,\frac{c}{C+F\cos(2N\theta-\phi)}\,\right]\label{regflow},
\eqn where $A$, $B$, $C$, $F$, $\phi$ are
\[
A=-\sinh^{2}2\chi+\frac{1}{2}\cos(2kD-\delta)\sinh4\chi,
\]
\[
B=-\sin(2kD-\delta)\sinh2\chi,\;\;C=1-A,
\]
\beq
F=\sqrt{A^{2}+B^{2}},\;\;\phi=\tan^{-1}(\frac{B}{A}).\label{abcf}
\eeq

It is well-known that in a random medium waves are always
localized in space and the localization is characterized by the
Lyapunov exponent (LE) $\gamma$. The Lyapunov exponent is defined
as \beq \gamma= \lim_{N \rightarrow \infty}
\langle\gamma_N\rangle, \label{lyap} \eeq with \beq \gamma_N
\equiv \frac{1}{2N}\ln (\frac{1}{T_{N}}). \eeq The fluctuation or
variance of $\gamma$ defined by \beq
\mbox{var}(\gamma)=\lim_{N\rightarrow\infty}
(\langle\gamma_{N}^{2}\rangle-\langle\gamma_{N}\rangle^2)\label{varlyap}
\eeq is a quantity which as will be shown below gives important
information about the system.

\subsection{Energy distribution in 1D systems}

The time averaged energy density ${\cal E}(x)$ at $x$ are defined
by \beq {\cal E}_{m}(x) = \frac{\rho_{m}}{4}
\left[\,|u_{m}|^{2}+\frac{|p_{m}|^{2}}{\rho_{m}^{2}c_{m}^{2}}\,\right].
\eeq By direct calculations we find that energy density in our
1D system is piece-wise constant and hence we can suppress the
redundant variable $x$. Energy density in $j$-th water block
${\cal E}^{w}_{j}$ and in $j$-th air block ${\cal E}^{g}_{j}$ are
given by \beq {\cal
E}^{w}_{j}=\frac{1}{2\rho\,c^{2}}\left[\,|W^{1}_{j}|^2+|W^{2}_{j}|^2\,\right]\label{ew}
\eeq and \beq {\cal
E}^{g}_{j}=\frac{1}{2\rho_{g}c_{g}^{2}}\left[\,|G^{1}_{j}|^2+|G^{2}_{j}|^2\,\right].\label{eg}
\eeq

In the regular case Eq.~(\ref{ew}) and (\ref{eg}) become \beq
{\cal E}^{w}_{j}=\frac{|f|^{2}}{2\rho
c^{2}}\left[\,M^{\dag}[(N-j)\theta]M[(N-j)\theta]\,\right]_{11}\label{ew1}
\eeq and \bqn {\cal E}^{g}_{j}&=&\frac{|f|^{2}}{2\rho_{g}
c^{2}_{g}}[\,M^{\dag}[(N-j+1)\theta](J^{-1})^{\dag}\non\\
&&\hspace{15mm}\times J^{-1}M[(N-j+1)\theta]\,]_{11},
\label{eg1}\eqn where

\bqn
f&=&\frac{\rho
cv}{{\cal M}_{11}\,e^{-ikD}-{\cal M}^{*}_{12}\,e^{ikD}}\nonumber\\
&=&\frac{\rho
cv}{\alpha(N\theta)\,e^{-ikD}-\beta^{*}(N\theta)\,e^{ikD}}.
\eqn

After lengthy but straightforward calculations Eq.~(\ref{ew1}) and
(\ref{eg1}) can be further simplified to \beq {\cal
E}^{w}_{j}=A_{w}+B_{w}\cos[2(N-j)\theta]\label{ew2} \eeq and \bqn
{\cal E}^{g}_{j} &=& C_{g}+A_{g}\cos[2(N-j+1)\theta]\non\\
 && +B_{g}\sin[2(N-j+1)\theta]\non\\
 &=& C_{g}+D_{g}\cos[2(N-j+1)\theta-\phi_{g}],\label{eg2}
\eqn where $A_{w}$, $B_{w}$, $A_{g}$, $B_{g}$, $C_{g}$, $D_{g}$
and $\phi_{g}$ are defined by \beq A_{w}=\frac{\rho
v^{2}}{2[C+F\cos(2N\theta-\phi)]} \cosh^{2}2\chi, \eeq \beq
B_{w}=-\frac{\rho v^{2}}{2[C+F\cos(2N\theta-\phi)]}
\sinh^{2}2\chi\hspace{15mm} \eeq
\[
A_{g}=\frac{\rho v^{2}}{4h[C+F\cos(2N\theta-\phi)]}\hspace{3cm}
\]
\beq
\times\left[\cosh 2\eta(1-\cosh 4\chi )+\cos\delta\sinh
2\eta \sinh 4\chi \right],
\eeq
\beq
B_{g}=\frac{\rho v^{2}\sin\delta\sinh 2\eta \sinh 2\chi}
{2h[C+F\cos(2N\theta-\phi)]},
\eeq
\[
C_{g}=\frac{\rho
v^{2}}{4h[C+F\cos(2N\theta-\phi)]}\hspace{3cm},
\]
\beq \times\left[\cosh 2\eta(1+\cosh 4\chi )-\cos\delta\sinh 2\eta
\sinh 4\chi\right], \eeq \beq
\phi_{g}=\tan^{-1}(\frac{B_{g}}{A_{g}}),\;\;\;\;
D_{g}=\sqrt{A^{2}_{g}+B^{2}_{g}}. \eeq The variables $C$, $F$,
$\phi$ in these equations are the same as in Eq.~(\ref{abcf}).

\subsection{Energy localization and collective behavior}

When waves propagate through media alternated with different
material compositions, multiple scattering of waves is established
by an infinite recursive pattern of rescattering. Writing
$p(x)=A(x)e^{i\theta(x)}$ with $A(x)=|p(x)|$ and $\theta$ being
the amplitude and phase respectively, the energy flow ${\cal J}
\sim \mbox{Re}[i (p^{*}(x)\partial_{x} p(x)]$ becomes ${\cal J}
\sim A^{2}\partial_{x}{\theta}$. Obviously, the energy flow will
come to a complete halt and the waves are localized in space when
$A$ does not equal to zero but phase $\theta$ is constant at least
by domains.

From these observations, we propose to use the phase behavior of
waves to characterize the wave localization. Expressing $p(x)$ and
$u(x)$ as \beq p(x)\equiv A_{p}(x)e^{i\theta_{p}(x)}\eeq and \beq
u(x)\equiv A_{u}(x)e^{i\theta_{u}(x)}. \eeq we construct two unit
phase vectors as \beq
\vec{v}_{p}\equiv\cos\theta_{p}\hat{e}_{x}+\sin\theta_{p}\hat{e}_{y}\label{vp}
\eeq and \beq
\vec{v}_{u}\equiv\cos\theta_{u}\hat{e}_{x}+\sin\theta_{u}\hat{e}_{y}.\label{vu}
\eeq Physically, these phase vectors represent the oscillation
behavior of the system. We can plot the phase vectors in a
two-dimensional plane.

\section{Numerical results and Discussion}

\subsection{Ordered cases}

In the ordered case, the interference of multiply scattered waves
leads to frequency band structures. For frequencies located in
pass bands, waves propagate through the whole system, while for
frequencies within a band gap, waves are evanescent.

\begin{figure}[hbt]
\epsfxsize=3in \epsffile{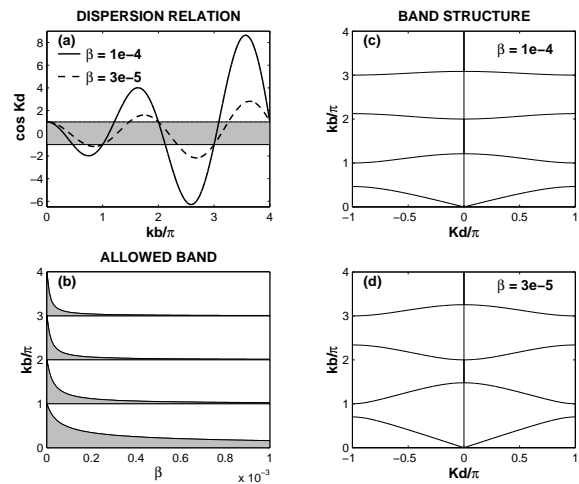}\vspace{5mm}
\caption{\label{figure2}\small Various properties of periodic
system. (a) Dispersion relations for air fractions $\beta=10^{-4}$
(solid line) and $\beta=3 \times 10^{-5}$ (broken line). Segments
of curves inside the gray region bounded by $\cos Kd=\pm 1$
correspond to the propagating waves (allowed bands), whereas those
outside of gray region represent evanescent wave. (band gaps). (b)
Allowed bands (gray regions) vs air fraction $\beta$. (c) Band
structure for $\beta=10^{-4}$. (d) Band structure for
$\beta=3\times 10^{-5}$. Here $K$ is the Bloch wave number, $d$ is
the lattice spacing.}
\end{figure}

In Fig.~\ref{figure2}(a), $\cos Kd$ versus $kb/\pi$ from
Eq.~(\ref{disp}) are displayed for $\beta=10^{-4}$(solid curve)
and $\beta=3\times 10^{-5}$ (broken curve). The segments of curves
between $\cos Kd=\pm 1$ (the gray region) give real solutions of
$Kd$ and correspond to pass bands, whereas those outside of gray
region correspond to forbidden bands. Fig.~\ref{figure2}(b) shows
the relation between the pass/forbidden bands and the fraction of
air blocks ($\beta$). It is seen that for $\beta > 10^{-3}$ and
$kb>3\pi$, the pass bands almost vanish. The band structures
for various bands are shown in
Fig.~\ref{figure2}(c) and (d). We see that as the volume fraction
$\beta$ increases, the width of the pass bands decrease and the
width of band gaps become larger. As expected from the earlier
discussion, that air blocks are very strong acoustic scatterer in
water leads to very wide band gaps and narrow pass bands shown in
Fig.~\ref{figure2}(c) and (d).

\begin{figure}[tbh] \epsfxsize=3in \epsffile{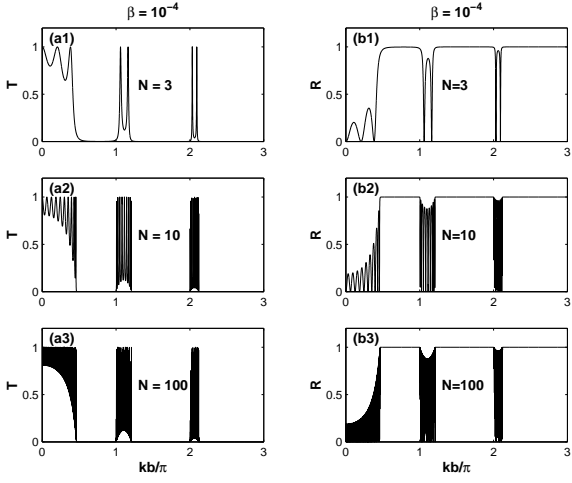}
\vspace{5mm}\caption{\label{figure3}\small (a1)-(a3) Transmission
rate versus frequency for $N=3,\,10,\,100$. (b1)-(b3) Reflection
rate versus frequency for $N=3,\,10,\,100$. The volume fraction of
air is $\beta=10^{-4}$.}
\end{figure}

\begin{figure}[tbh]
\epsfxsize=3in \epsffile{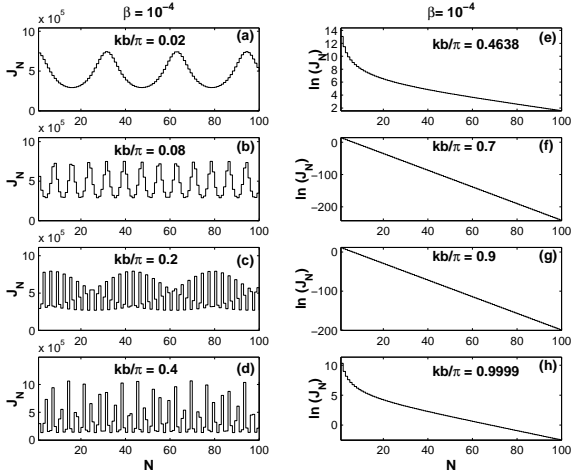}
\vspace{5mm}\caption{\label{figure4}\small (a)-(d) Energy flow vs
number of air blocks $N$, four frequencies are chosen from first
allowed band. (e)-(h) Log-Energy flow versus $N$ in first band
gap. The volum fraction of air is $\beta=10^{-4}$.}
\end{figure}

In Fig.~\ref{figure3}, transmission and reflection versus
frequency for different $N$ are plotted according to
Eqs.~(\ref{regtn}) and (\ref{regrn}). Inside a pass band,
transmission is significant. The transmission rate $T_{N}$
oscillates as the frequency varies, and it has $N-1$ peaks as
expected from Eq.~(\ref{regtn}). At these peaks $\sin N\theta=0$,
$T_{N}=1$ and $R_{N}=0$. That is, at the frequencies waves tunnel
through the whole system completely, a phenomenon of resonant
tunneling. Within a band gap, on contrast, although it is not
exactly zero due to the finiteness of the system considered, the
transmission is significantly prohibited. We have approximately
$T_{N}\approx 0$ and $R_{N}\approx 1$. Using Eq.~(\ref{regflow}),
energy flow ${\cal J}_{N}$ as function of the air-block number $N$
for different frequencies are plotted in Fig.~\ref{figure4}. The
frequencies appearing in diagrams (a), (b), (c), (d) are chosen
from the first pass band. These diagrams show that the flow varies
periodically as $N$ increases. When the frequency is low, for
instance at $kb/\pi=0.02$, a simple periodic behavior of ${\cal
J}_{N}$ with long space period is observed. As we increase the
frequency, the period becomes shorter, as illustrated in diagram
(b). If we further increase the frequency, the oscillation of
${\cal J}_{N}$ becomes more rapid and the period is smaller than
$d$. However, since $N$ is a integer, the behavior
of ${\cal J}_{N}$ tends to be more complicated as $N$ is enlarged
shown in diagrams (c) and (d).

When frequency enters into the band gap, waves become evanescent,
as indicated in diagrams (e), (f), (g), and (h). Diagram (e) and
(h) display the decay behavior of ${\cal J}_{N}$ at frequencies
very close to the gap edges. The behavior of ${\cal J}_{N}$ at
frequencies near the center of first band gap are illustrated in
diagrams (f) and (g), showing an exponential decay behavior. We
see that the decay rates in (f) and (g) are much more stronger
than in (e) and (h), as expected.

\begin{figure}[tbh]
\epsfxsize=3in\epsffile{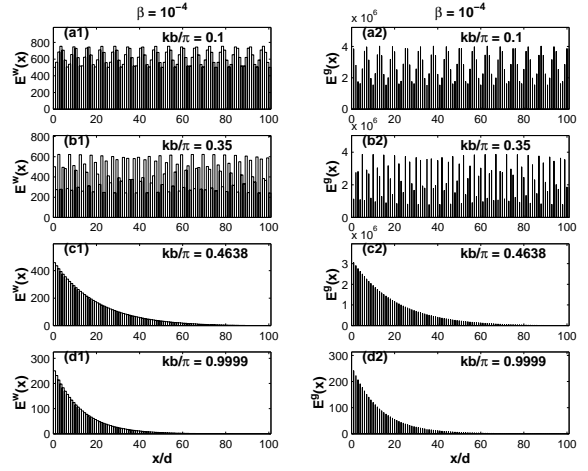}\vspace{5mm}
\caption{\label{figure5}\small Energy density distributions along
the duct at four different frequencies: $kb/\pi=0.02$, $0.35$,
$0.4638$, $0.9999$. Left (a1)-(d1): Energy density in water
blocks. Right (a2)-(d2): Energy density in air blocks. Number of
air blocks $N=100$ and air fraction $\beta=10^{-4}$.}
\end{figure}

Using Eq.~(\ref{ew2}) and (\ref{eg2}), energy density can be
calculated. Fig.~\ref{figure5} shows the energy density
distribution along the duct for $N=100$ air blocks. On the left
panel of Fig.~\ref{figure5}, i.~e. diagrams (a1), (b1), (c1), and
(d1), the energy density in water blocks are shown, whereas on the
right panel, i.~e. diagrams (a2), (b2), (c2), and (d2), the energy
density in the air blocks are displayed. In the computation, the
air fraction is taken as $\beta=10^{-4}$. We find that for
frequencies located in the pass bands, the energy density varies
periodically along the traveling path. For frequencies within the
frequency gaps, the energy is trapped near the acoustic source,
and the energy density decays about exponentially along the path.
Meanwhile, the energy flow is calculated to be close to zero.

\begin{figure}[tbh]
\epsfxsize=3in\epsffile{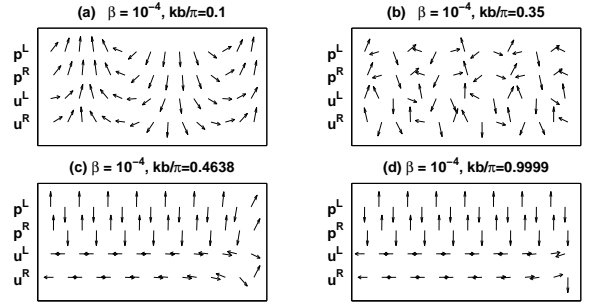}\vspace{5mm}
\caption{\label{figure6}\small Phase vectors at air block
boundaries. Total number of air blocks is $N=15$, and the air
fraction is $\beta=10^{-4}$. Four frequencies are chosen according
to Fig.~\ref{figure5}, that is, $kb/\pi=0.02$, and $0.35$ in the
first allow band and $0.4638$, $0.9999$ in the first gap. In the
diagrams, the small arrows are used to represent the phase
vectors.}
\end{figure}

We now look at the oscillation behavior of the blocks. As
discussed above, zero energy flow leads to a phase coherence
behavior of the medium. For the purpose we consider the behavior
of the phase vectors $\vec{v}_{p}$ and $\vec{v}_{u}$ defined in
Eqs.~(\ref{vp}) and (\ref{vu}). Typical results of phase
behavior are shown in Fig.~\ref{figure6}. For further convenience, only
the phase vectors at the interfaces between air and water are
shown. Symbols $p^{L}$, $p^{R}$, $u^{L}$, $u^{R}$ appearing in
Fig.~\ref{figure6} denote respectively the phase vectors for the
pressure and the velocity fields on the left and right side of the
air blocks. We set the vibration phase of LB to $1$. First we
discuss the cases shown by diagrams (a) and (b). The frequencies
in (a) and (b) are chosen from the first pass band. The phase
vectors in these two cases point to various directions along the
duct and $\vec{v}_{p}\cdot \vec{v}_{u}\neq 0$, resulting in a
nonzero energy flow. The waves are extended in the system. In
diagrams (c) and (d), the frequencies are chosen from the first
band gap in which waves are evanescent as shown in
Fig.~\ref{figure5}. In this case, all the phase vectors of the
pressure field are pointing to either $\pi/2$ or $-\pi/2$, and are
perpendicular to the phase vectors of the velocity field. The
pressure at the two sides of any air block is almost in phase.
Different from the higher dimensional cases in which all phase
vectors of localized fields point to the same
direction\cite{Ye2,Ye}, the present phase vectors are constant
only by domains. The velocity field in neighboring domains
oscillate with a phase difference $\pi$. At the far end of the
sample, however, the phase vectors become gradually disoriented,
implying that the energy can leak out only at the boundary due to
the finite sample size. We note here that such a phase ordering
not only exists for the boundaries of the air blocks, but also
appears inside the whole medium.

\subsection{Disordered situations}

Unlike in the ordered case for which waves can propagate through
all air blocks if the frequency is located in pass band, in a
disordered 1D system waves are always localized. In this section,
we study numerically acoustic propagation in the disordered case.

\begin{figure}[tbh]
\epsfxsize=3in\epsffile{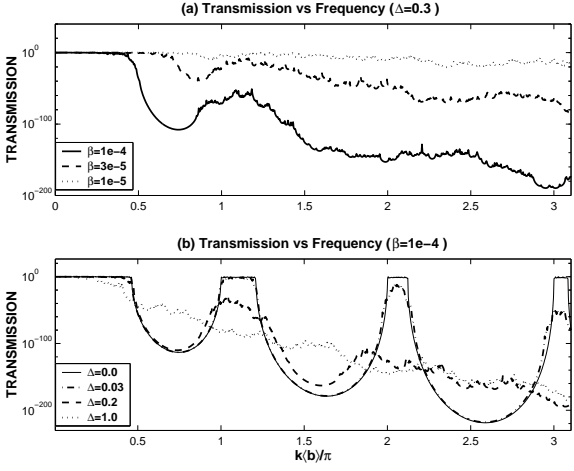}
\vspace{10pt}\caption{\label{figure7}\small Transmission versus
$k\langle b\rangle/\pi$ for various air fractions at $\Delta =
0.3$ (a) and different disorders (b). The number of the air-blocks
is 100.}
\end{figure}

Fig.~\ref{figure7}(a) presents the typical results of the
transmission rate as a function of $k\langle b\rangle$ for various
$\beta$ at a given randomness.  At frequencies for which the
wavelength is smaller than the averaged distance between air
blocks, the transmission is significantly reduced by increasing
air fraction.

Fig.~\ref{figure7}(b) illustrates the effect of the randomness
$\Delta$ on transmission for a given air-fraction. For comparison,
the transmission in the corresponding regular array ($\Delta = 0$)
is also plotted. The gaps are located between $k\langle
b\rangle/\pi = 0.4638$ and $1$, $1.21$ and $2$, $2.128$ and $3$, and
so on. We find that for frequencies located inside the band gaps
of the corresponding periodic array, the disorder-induced
localization effect competes yet reduces the band gap effect. To
characterize wave localization in this case, both the band gap and
the disorder effects should be considered, supporting the two
parameter scaling theory\cite{Deych}. However, increasing disorder
tends to smear out the band structures. When exceeding a certain
amount, the effect from the disorder suppresses the band gap
effect completely, and there is no distinction between the
localization at frequencies within and outside the band gaps.

Fig.~\ref{figure7} shows that the localization behaviour depends
crucially on whether the wavelength ($\lambda$) is greater than
the average distance between air blocks. When $\langle
b\rangle/\lambda$ is less than $1/4$, the localization effect is
weak. We also observe that with the added disorder, the
transmission is enhanced in the middle of the gaps. Similar
enhancement due to disorder has also been reported
recently\cite{Maradudin95}. Differing from \cite{Maradudin95},
however, the transmission at frequencies within the gaps of the
corresponding periodic arrays in the present system is not always
enhanced by disorder. Instead the transmission is reduced further
by the disorder near the band edges.

\begin{figure}[tbh]
\epsfxsize=3.2in\epsffile{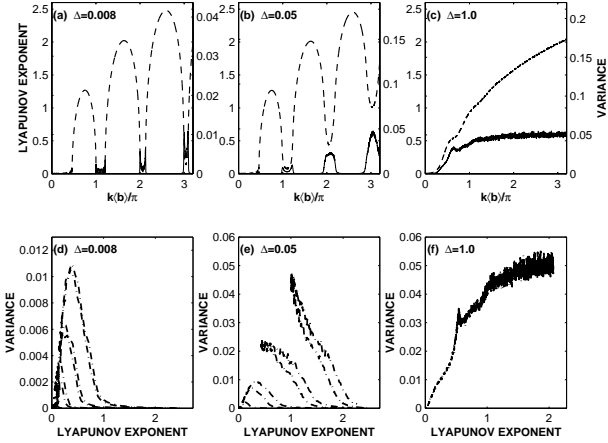}
\vspace{10pt}\caption{\label{figure8}\small Diagrams (a), (b), and
(c) show the Lyapunov exponent (LE) in broken lines and its
variance in solid lines as a function of $k\langle b\rangle/\pi$
for three random situations. Diagrams (d), (e), and (f) present
the plots of the exponent versus its variance at the three random
cases. Here $\beta=10^{-4}$.}
\end{figure}

To further explore the transmission property, we calculate
Lyapunov exponent $\gamma$ and its variance $\mbox{var}(\gamma)$
according to Eq.~(\ref{lyap}) and (\ref{varlyap}). The sample size
is chosen in such a way that that it is much larger than the
localization length and the ensemble average is carried out over
$2000$ random configurations. Fig.~\ref{figure8} presents the
results for Lyapunov exponent (LE) and its variance as a function
of $k\langle b\rangle /\pi$ for various randomness. As expected,
when the randomness is small LE mimics the band structures and the
variance of LE inside the gaps is small. Contrast to the optical
case\cite{Deych}, there are no double maxima for the variance
inside the gap. Rather, the double peaks appear in the the allow
bands when the system is disordered. The double peak feature is
more prominent in the low frequency bands. When the randomness
exceeds a certain value, however, the double peaks emerge. The
higher frequency, the lower is the critical value. For example,
the double peaks are still visible in the first allow band (c.~f.
Fig.~\ref{figure8}(b)), while there is only one peak inside the
higher pass bands. Meanwhile, the increasing disorder reduces the
band gap effect and smears out the oscillation in LE, in
accordance with Fig.~\ref{figure8}. We also plot LE versus its
variance in Fig.~\ref{figure8}. When randomness is weak, several
branches appear in the LE-variance relation. The frequency range
of a branch covers a pass band. In the optical case, the frequency
range of a branch covers a gap instead\cite{Deych}. A prominent
feature in the present system is that when the double peaks in the
variance are destroyed, the minima of LE correspond to the maxima
of its variance. This is different from the optical
case\cite{Deych}.

With increasing disorder, we do not observe the genuine linear
dependence between LE and its variance, as expected from the
single parameter scaling theory\cite{TAA}, indicating that the
single parameter scaling law may not be applicable in the present
system.

\begin{figure}[tbh]
\epsfxsize=3in\epsffile{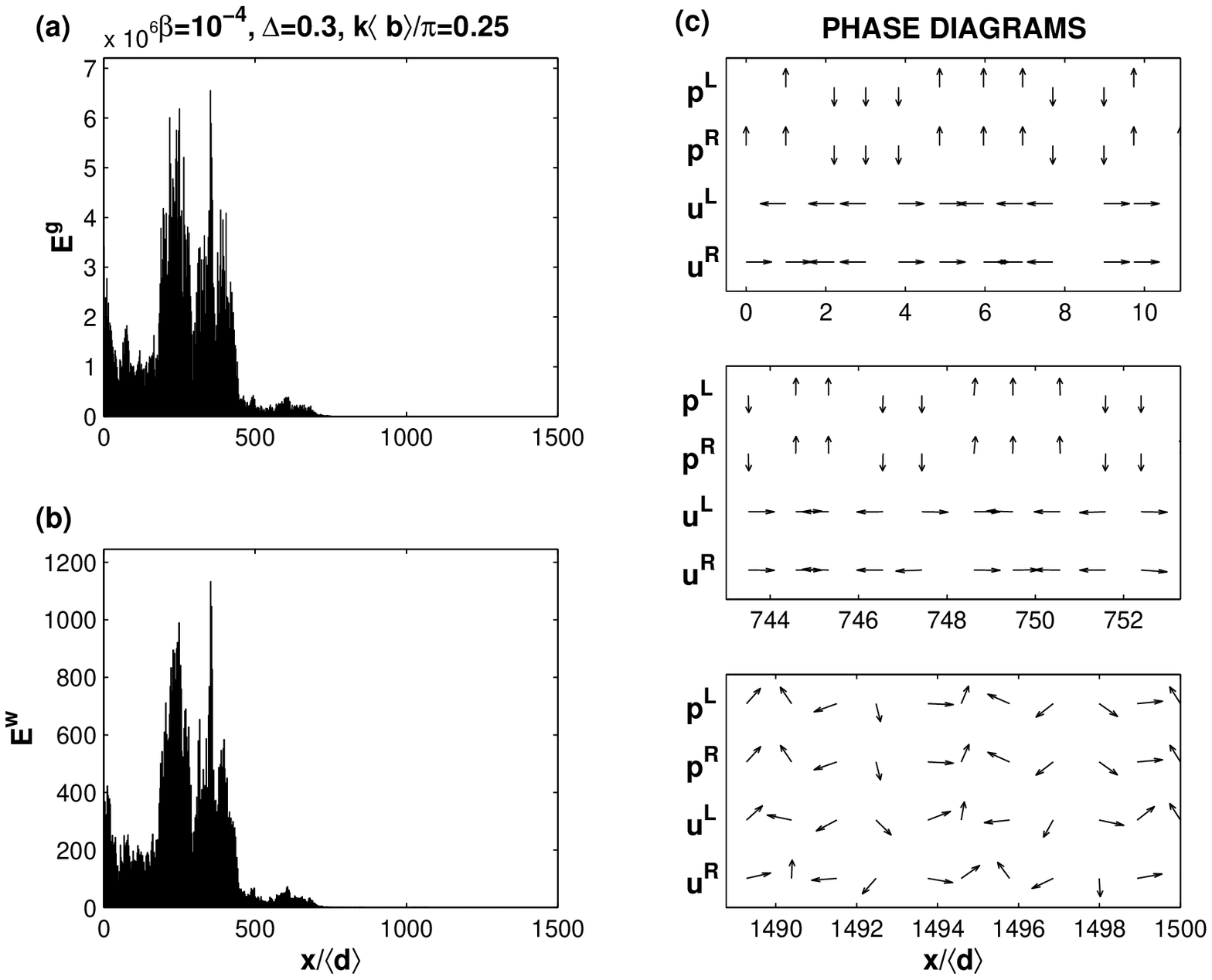}\vspace{5mm}
\caption{\label{figure9}\small Energy density distributions along
the duct in air blocks (a) and in water blocks (b). Phase vectors
at the interfaces for three spatial ranges of the medium are
illustrated in (c). The unit of energy density is $J/m^{3}$, air
fraction $\beta=10^{-4}$, $k\langle b\rangle/\pi=0.25$,
$\Delta=0.3$, $N=1500$.}
\end{figure}

\begin{figure}[tbh]
\epsfxsize=3in\epsffile{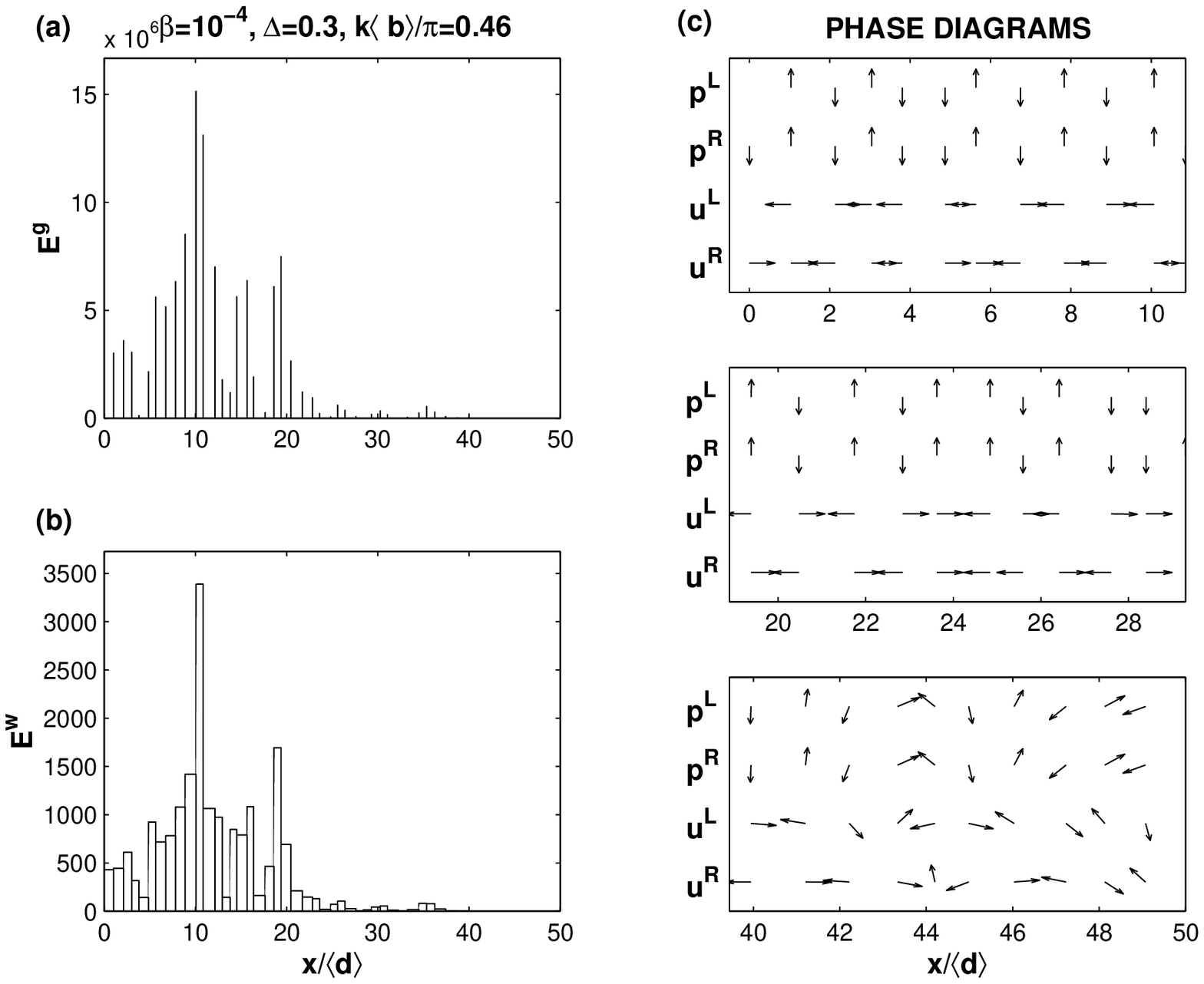}\vspace{5mm}
\caption{\label{figure10}\small Energy density distributions along
the duct in air blocks (a) and in water blocks (b). Phase vectors
at the interfaces for three spatial ranges of the medium are
illustrated in (c). The unit of energy density is $J/m^{3}$, air
fraction $\beta=10^{-4}$, $k\langle b\rangle/\pi=0.46$,
$\Delta=0.3$, $N=50$.}
\end{figure}

\begin{figure}[tbh]
\epsfxsize=3in\epsffile{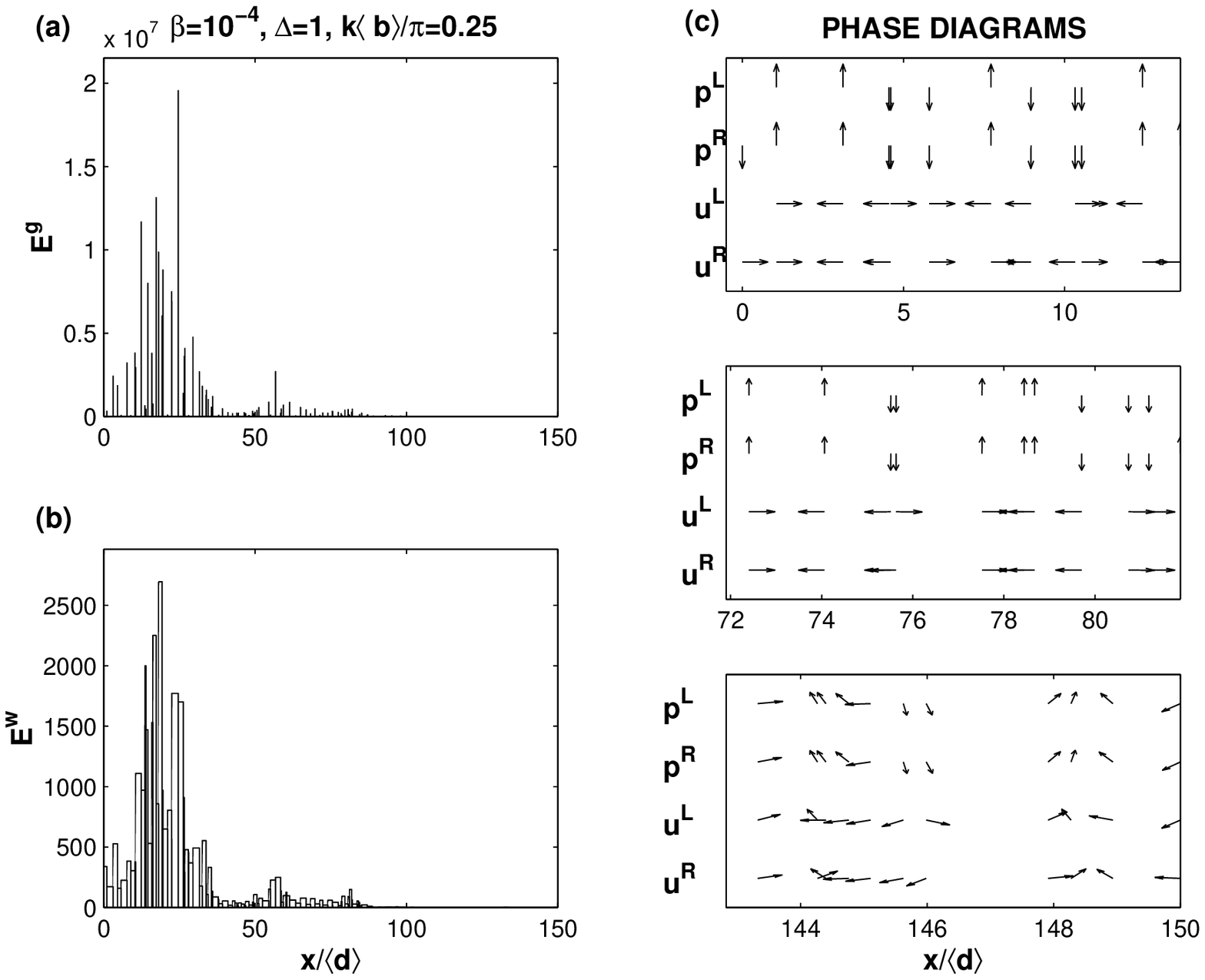}\vspace{5mm}
\caption{\label{figure11}\small Energy density distributions along
the duct in air blocks (a) and in water blocks (b). Phase vectors
at the interfaces for three spatial ranges of the medium are
illustrated in (c). The unit of energy density is $J/m^{3}$, air
fraction $\beta=10^{-4}$, $k\langle b\rangle/\pi=0.25$,
$\Delta=1$, $N=150$.}
\end{figure}

Eqs.~(\ref{ew}) and (\ref{eg}) provide the formulas for studying
energy distribution in any situation. As discussed in the regular
case, when wave is localized, a kind of ordering of the phase
vectors appears. From the alignment pattern of phase vectors we
can know whether waves inside the system are localized. For air
fraction $\beta=10^{-4}$, three different cases with different
$k\langle b\rangle/\pi$, $\Delta$ and $N$ are shown in
Fig.~\ref{figure9}, \ref{figure10},\ref{figure11}. To isolate the
localization effect from the band gap effects, we choose the two
frequencies located in the first allow band: one is in the middle
of the band and the other is near the lower band edge.

First, we note that the energy density is constant in each
individual block. This is a special feature of 1D classical
systems, and can be verified by a deduction from Eqs.~(\ref{ew})
and (\ref{eg}). From these figures, we observe that when the
sample size is sufficiently large, waves are always localized for
any given amount of randomness. When localized, the waves are
trapped inside the medium, but not necessarily confined at the
site of the source, unless the band gap effect is dominant. The
energy distribution does not follow an exponential decay along the
path. This differs from situations in higher
dimensions\cite{Ye1,Ye}. It is also shown that the energy stored
in the medium can be tremendous.

These figures also show that for low frequencies and when the
randomness is small, to trap the waves a large number of air blocks
is needed. Like in the regular cases, when waves are
localized, the coherent behavior of the medium appears, and the
phases are constant by domains. The phase vector domains are
sensitive to the arrangement of the air blocks. Moreover, when
localization is evident, increasing the sample size by adding more
air blocks to the far end of the system will not change the
patterns of the energy distribution and phase vectors. Therefore
the energy localization and the phase coherence behavior are not
caused by the boundary effect.

Fig.~\ref{figure9} shows the energy distribution inside the
medium. Here we choose $k\langle b\rangle/\pi=0.25$, which is
inside first pass band. We see that the energies are localized in
both air and water blocks. However, unlike in the regular cases,
the energy does not always decay exponentially along the path. In
Fig.~\ref{figure10} we use the same parameters as in
Fig.~\ref{figure9} except that we choose $k\langle
b\rangle/\pi=0.46$, which is very close to the gap edge $0.4638$.
We see now that waves can be very easily trapped by using only
$N=50$ air blocks. In Fig.~\ref{figure11} we choose  $k\langle
b\rangle/\pi=0.25$ as used in Fig.~\ref{figure9} but increase the
randomness to $\Delta=1$, waves are trapped by $N=150$ air blocks,
fewer than in the case of Fig.~\ref{figure9}.

\section{Conluding Remarks}

In this paper we studied the propagation of acoustic waves in 1D
layered system consisting of air and water blocks. Both regular
and random arrangements of air blocks have been studied. We first
derived the basic formulas for the general situations, and then
applied these formulas to regular and random cases separately. For
periodically placed air blocks, the band structures were studied
and were shown to have large band gaps. The transmission,
reflection, Luapunov exponent and its variance, energy
distribution, energy flow and medium vibration were also studied.
For the case of randomly placed air blocks, the results pointed
out that waves are always confined in a finite spatial region. The
disorder leads a significant energy storage in the system. It is
also indicated that the wave localization is related to a
collective behavior of the system in the presence of multiple
scattering, also observed for higher dimensions\cite{Ye2,Ye}. The
appearance of such a collective phenomenon may be regarded as an
indication of a kind of classical Goldstone modes in the context
of the field theory\cite{Umezawa}.

\section*{Acknowledgments}

The work received support from National Science Council (No.
NSC89-2611-M008-002 and NSC89-2112-M008-008).

\newpage
\newpage
\section*{Figures}

\input{epsf}

\begin{center}

\bf Figure 1

\end{center}

\begin{center}

\bf Figure 2

\end{center}

\newpage

\begin{center}

\bf Figure 3

\end{center}

\begin{center}

\bf Figure 4

\end{center}

\newpage

\begin{center}

\bf Figure 5

\end{center}

\begin{center}

\bf Figure 6

\end{center}

\newpage

\begin{center}

\bf Figure 7

\end{center}

\begin{center}

\bf Figure 8

\end{center}

\newpage

\begin{center}

\bf Figure 9

\end{center}

\begin{center}

\bf Figure 10

\end{center}

\newpage

\begin{center}

\bf Figure 11

\end{center}

\end{document}